\documentclass[%
 reprint,
 groupedaddress,
superscriptaddress,
 amsmath,amssymb,
 aps,
pra,
]{revtex4-1}

\usepackage{graphicx}% Include figure files
\usepackage{dcolumn}% Align table columns on decimal point
\usepackage{bm}% bold math
\begin{document}

\preprint{APS/123-QED}

\title{Bounding the outcome of a two-photon interference measurement using weak coherent states}% Force line breaks with \\

\author{Andr\'{e}s Aragoneses}
\affiliation{Department of Physics, Duke University, Durham, North Carolina 27708, USA}
\affiliation{Department of Physics and Astronomy, Carleton College, Northfield, Minnesota 55057, USA}
\author{Nurul T. Islam}%
\affiliation{Department of Physics, Duke University, Durham, North Carolina 27708, USA}

\affiliation{Department of Physics, The Ohio State University, 191 West Woodruff Ave., Columbus, Ohio 43210 USA}

\author{Michael Eggleston}
\affiliation{Department of Physics, Duke University, Durham, North Carolina 27708, USA}

\author{Arturo Lezama}
\affiliation{Department of Physics, Duke University, Durham, North Carolina 27708, USA}
\affiliation{Instituto de F\'isica, Universidad de la Rep\'ublica, C.P. 30, 11000 Montevideo, Uruguay}
\author{Jungsang Kim}
\affiliation{Department of Electrical Engineering and the Fitzpatrick Institute for Photonics, Duke University, Durham, North Carolina 27708, USA}

\author{Daniel J. Gauthier}
\email[corresponding author:~]{gauthier.51@osu.edu}
\affiliation{Department of Physics, The Ohio State University, 191 West Woodruff Ave., Columbus, Ohio 43210 USA}

\date{\today}% It is always \today, today,

\begin{abstract}
Interference of two photons at a beamsplitter is at the core of many quantum photonic technologies, such as quantum key distribution or linear-optics quantum computing. Observing high-visibility interference is challenging because of the difficulty of realizing indistinguishable single-photon sources. Here, we perform a two-photon interference experiment using phase-randomized weak coherent states with different mean photon numbers.  We place a tight upper bound on the expected coincidences for the case when the incident wavepackets contain single photons, allowing us to observe the Hong-Ou-Mandel effect. We find that the interference visibility is at least as large as 0.995$^{+0.005}_{-0.013}$. 
\end{abstract}
\maketitle
\begin{figure}[htb]
   \centering
     \resizebox{0.75\columnwidth}{!}{\includegraphics{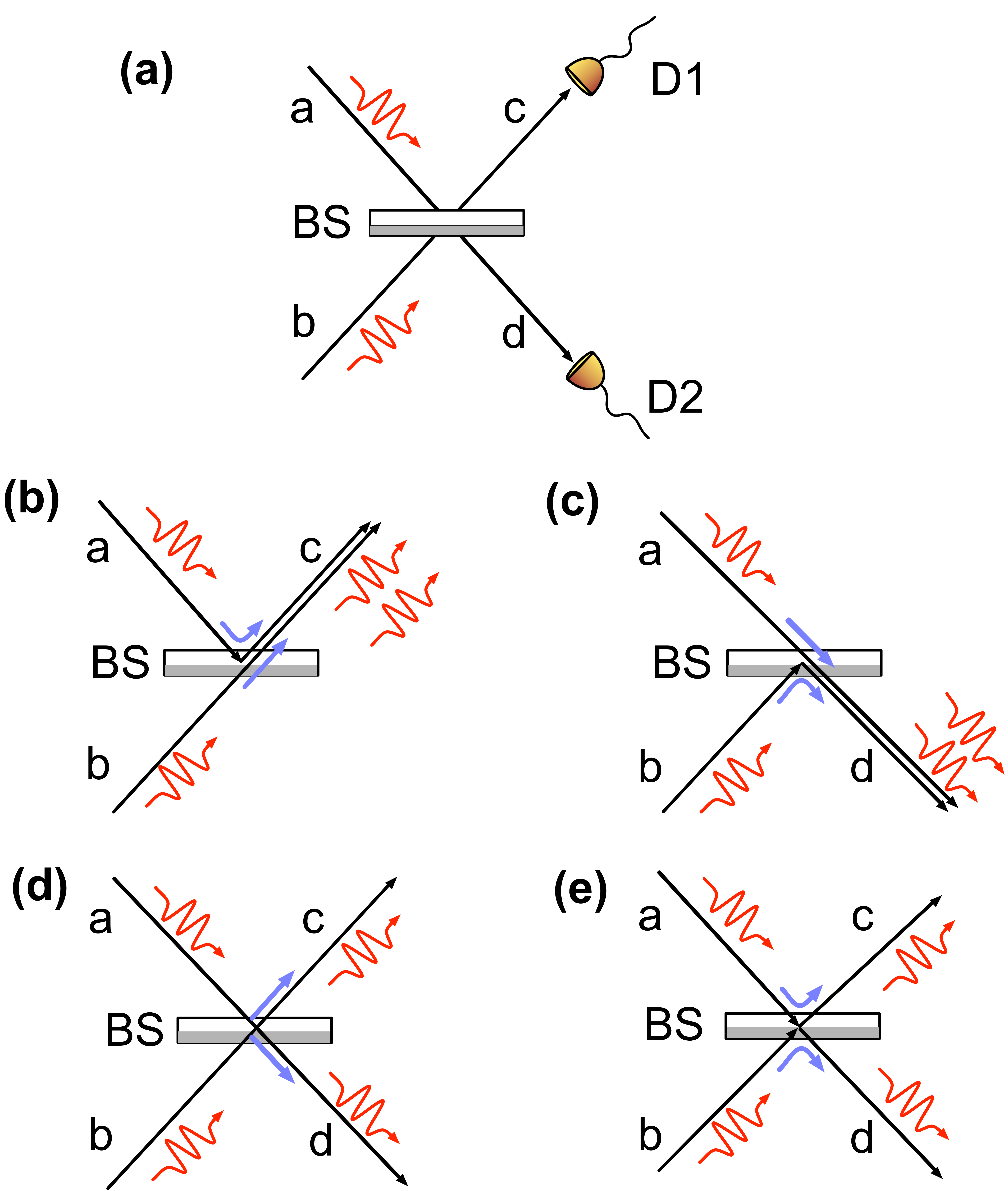}}
    \caption{(a) Hong-Ou-Mandel interference at a beamplitter. (b)-(e) Illustration of the four quantum mechanical pathways for photon interaction at the beamsplitter.  The Hong-Ou-Mandel effect arises from destructive interference between the pathways (d) and (e). \label{fig1}}
\end{figure}

One central effect in quantum optics is the interference of two indistinguishable single-photon wavepackets at a beam splitter, as illustrated in Fig. \ref{fig1}(a).  When single-photon counting detectors are placed in the output port, quantum mechanics predicts that there are no coincident events in the ideal situation, indicating that there are never single photons appearing in each output port. The only possibilities are two photons emerging from one output port or the other (Figs. \ref{fig1}(b) and (c)).

This fusion or bunching of photons, known as the Hong-Ou-Mandel (HOM) effect and observed over 30 years ago \cite{1987_PRL_HOM, Walmsley-2017}, arises from the destructive interference between the two quantum mechanical probability amplitudes for single photons emerging in each of the two output ports (Figs. \ref{fig1}(d) and (e)), and highlights the quantum nature of light.  The two-photon visibility $V=1-g^{(2)}(0)$ is equal to 1 in this case, where $g^{(2)}(0)$ is the minima of the normalized second-order (photon-photon) coherence function.

While the technology for generating indistinguishable single-photon wavepackets has progressed rapidly over time, it is highly desirable to use simpler sources, such as attenuated laser pulses.  Unfortunately, when the single-photon wavepackets are replaced by phase-randomized weak coherent-state (PRWCS) wavepackets, which can be generated with attenuated lasers, HOM interference is obscured by the presence of multi-photon wavepackets as governed by Poisson statistics.  For these highly classical states, $V$=1/2 in the ideal case \cite{1983_PRA_Mandel,2005_JOB_Loudon,2013_Nat_Comm_Jeongwan,2016_CPB_Chen,2017_QST_Tittel,2018-Bing}.  
Here, based on the proposals and analyses of Yuan \textit{et al.} \cite{Yuan-2016} and Navarrete \textit{et al.} \cite{2017_Arxiv_Curty}, we show that it is possible to place a tight upper bound on the outcome of the two-photon interference experiment using PRWCS's if we perform measurements with different mean photon numbers, as discussed in detail below.  In particular, we upper-bound the expected coincidence probability for one photon emerging from each output port of the beam splitter conditioned on the presence of two incident single-photon wavepackets $P(1,1|1,1)$, even though we use wavepackets with a fluctuating photon number.  We stress that no post-selection of events is used in the analysis; all measurements are integrated into the expression for the upper bound.  In related work, Valente and Lezama \cite{2017_JOSAB_Lezama} recently performed quantum tomography of single-photon temporal states using PRWCS with varying mean photon numbers.  

To see how experiments using these states can be used to bound the outcome of a two-photon interference experiment, we used the analysis presented in Ref.~\cite{2017_Arxiv_Curty} for the case of a two-mode input and two-mode output photonic circuit appropriate for the setup illustrated in Fig. \ref{fig1}(a). For clarity of presentation, we first consider the case that the detectors have no dark counts, then describe how to account for this non-ideality below. Briefly, the probability for a measured coincidence (mnemonic $C$) at the output ports is given by
\begin{align}
C^{\mu_a,\mu_b}=&\kappa_1\kappa_2\mu_a\mu_bP(1,1|1,1)+\kappa_1\kappa_2\frac{\mu_a^2}{2}P(1,1|2,0) \nonumber \\
&+\kappa_1\kappa_2\frac{\mu_b^2}{2}P(1,1|0,2)+\mathcal{O}(\mu_a^r,\mu_b^s,\mu_a^t\mu_b^u).
\label{Pmumucc}
\end{align}
Here, $P(n_c,n_d|n_a,n_b)$ is the probability of $n_c$ and $n_d$ photons in the output ports $c$ and $d$, respectively, conditioned on the presence of $n_a$ and $n_b$ photons in input ports $a$ and $b$, respectively, $\mu_i$ is the mean photon number of the PRWCS for input port $i$, $\kappa_j$ is the efficiency of detector $j$, and $\mathcal{O}$ is a positive-definite quantity related to higher-order terms in the $\mu$'s with integer powers $r,s,t,u > 1$.

The first term on the right-hand-side of the equation represents the HOM effect and accounts for both of the scattering processes in Fig. \ref{fig1}(b) and (c); $P(1,1|1,1)$=0 using the predictions of quantum mechanics, although we make no assumption about the value for this probability.  The term proportional to $P(1,1|2,0)$ [$P(1,1|0,2)$] is due to a two-photon wavepacket in port $a$ [$b$] and an empty wavepacket in port $b$ [$a$]. For the case when $\mu_a=\mu_b$ these terms have similar-size coefficients regardless of the smallness of the mean photon number, which is the reason why $V$ is limited to a value of 1/2 or less.

The goal of the protocol described in Refs. \cite{Yuan-2016} and \cite{2017_Arxiv_Curty} is to isolate a desired conditional probability in Eq. \ref{Pmumucc}.  This is accomplished by noting that the higher-order terms in Eq. \ref{Pmumucc} fall off rapidly when the $\mu$'s are not too large. Thus, it is possible to truncate the higher-order terms at some point, leaving $N$ unknown conditional probabilities.  It is possible to determine these conditional probabilities by performing multiple experiments with $N$ total different values of $\mu_a$ and $\mu_b$~\footnote{In the quantum cryptography community, the largest values of $\mu_a$ and $\mu_b$ are often called `signal' states, whereas smaller values (including zero mean photon number) are often called `decoy' states.}.  

For the two-photon interference experiment considered here, we want to isolate $P(1,1|1,1)$, which only involves single photons in each port.  Thus, we focus on the case when $\mu_a,\mu_b \ll 1$.  Under this condition, the higher-order terms represented by $\mathcal{O}$ in Eq. \ref{Pmumucc} will contribute negligibly to $C^{\mu_a,\mu_b}$ and will be dropped when we obtain an upper bound on $P(1,1|1,1)$ below.  There are only three lowest-order conditional probabilities to determine, indicating that only three different experiments are required.

To this end, we consider using the following set of mean photon numbers: 1) States with $\mu_a=\mu_b=\mu$; and two other states with 2) port $a$ blocked so that $\mu_a=0$, $\mu_b=\mu$, and with 3) port $b$ blocked so that $\mu_a=\mu$ and $\mu_b=0$.  By inspecting Eq. \ref{Pmumucc}, we see that the each state with a blocked input port isolates one of the multi-photon terms so that
\begin{eqnarray}
C^{\mu,0}=\kappa_1\kappa_2\frac{\mu^2}{2}P(1,1|2,0)+\mathcal{O}_a(\mu^r), \label{Pmu0cc} \\
C^{0,\mu}=\kappa_1\kappa_2\frac{\mu^2}{2}P(1,1|0,2)+\mathcal{O}_b(\mu^s). \label{P0mucc}
\end{eqnarray}
Combining Eqs. \ref{Pmumucc}, \ref{Pmu0cc} and \ref{P0mucc}, we obtain
\begin{equation}
C^{\mu,\mu}-C^{\mu,0}-C^{0,\mu}=\kappa_1\kappa_2\mu^2P(1,1|1,1)+\mathcal{O}_{ab}(\mu^t\mu^u),
\label{PmumuHOM}
\end{equation}
where
\begin{equation}
\mathcal{O}_{ab}(\mu^t\mu^u)=\mathcal{O}(\mu^r,\mu^s,\mu^t\mu^u)-\mathcal{O}_a(\mu^r)-\mathcal{O}_b(\mu^s) \geq 0.
\end{equation}

An upper bound for the desired conditional probability, denoted by $P(1,1|1,1)^{ub}$, is obtained by using the fact that $\mathcal{O}_{ab}(\mu^t\mu^u) \geq 0$, dropping this term in Eq. \ref{PmumuHOM}, and solving for the bound.  We find that
\begin{equation}
P(1,1|1,1)^{ub}=\frac{C^{\mu,\mu}-C^{\mu,0}-C^{0,\mu}}{\kappa_1\kappa_2\mu^2} \geq P(1,1|1,1). \label{P11bound}
\end{equation}
If $\mu$ and $\kappa_i$ are well calibrated, measuring three coincidence count probabilities can be combined via Eq. \ref{P11bound} to reveal the HOM two-photon interference effect.

To avoid this calibration process with the aim of only using the measured counts, we seek to place a bound on the denominator appearing in Eq. \ref{P11bound}.  Considering the two decoy experiments, the single-count (mnemonic $S$) probabilities for each detector when photons are present in both input ports are given by
\begin{align}
S^{D1}&=\kappa_1\mu(P(1,0|1,0)+P(1,0|0,1))+\mathcal{O}_{Sc} \nonumber \\
&=\kappa_1\mu+\mathcal{O}_{Sc}, \\
S^{D2}&=\kappa_2\mu(P(0,1|1,0)+P(0,1|0,1))+\mathcal{O}_{Sd}\nonumber \\
&=\kappa_2\mu+\mathcal{O}_{Sd},
\end{align}
where $\mathcal{O}_{Sc},\mathcal{O}_{Sd} \geq 0$ are terms of quadratic order or higher in the $\mu$'s. Thus, we obtain the lower bound
\begin{equation}
\left(\kappa_1\kappa_2\mu^2\right)^{lb}=S^{D1}S^{D2}. \label{kappamu}
\end{equation}
Inserting Eq. \ref{kappamu} into Eq. \ref{P11bound} results in the upper bound
\begin{equation}
P(1,1|1,1)^{ub}=\frac{C^{\mu,\mu}-C^{\mu,0}-C^{0,\mu}}{S^{D1}S^{D2}}, \label{P11bound-cnts}
\end{equation}
which only depends on measured count statistics and does not require careful calibration of the $\mu$'s or $\kappa$'s.

Generalizing our results to account for detector dark counts involves an additional measurement where both inputs to the beam splitter are blocked and recording dark-count-induced coincidences and single counts.  We then subtract the appropriate dark events from the $C$'s and $S$'s.  Equation~\ref{P11bound-cnts} then applies using these corrected values for the $C$'s and $S$'s.

To test this approach experimentally, we generate photonic wavepackets using a highly-attenuated, gain-switched vertical cavity semiconductor laser (VCSEL, Vixar 680M-0000-X002) operating at 680 nm, and a repetition rate of either 3.91 or 31.25 MHz. The laser pulse has a temporal width $>$7 ps and is non-transform limited with chirp that varies from pulse-to-pulse. The beam generated by the laser is attenuated, coupled into a single-mode optical fiber, recollimated, sent through a linear polarizer, and split. One beam reflects from a mirror attached to a piezoelectric actuator driven by a ramp to randomize the relative phase, and passes through an adjustable optical delay element consisting of a corner cube mounted on a linear translation stage (Zaber, T-LSR150B). 

The two beams are combined on a nearly symmetric beam splitter with a nominal intensity reflection (transmission) coefficient of 0.52 (0.48). The precise value of $R=1-T$ is highly sensitive to the angle of incidence and can be high as $R=0.54$ ($T=0.46$) and as low as $R=0.50$ ($T=0.50$). The light emerging from each output port of the beam splitter is sent to single-photon counting detectors (Perkin-Elmer, SPCM-AQ4C, 60\% nominal detection efficiency, $<0.5\%$ afterpulsing probability, $<10^3$ dark counts per second).  We do not purposefully adjust the detector efficiency or optical loss from the beam splitter to the detectors, but we find that the overall efficiencies (optical loss times detector efficiency) for each path are the same to within $\pm$5\% based on measurements of the single count rates.  The electrical pulses generated by the detectors are sent to single- and coincident-event counters, which only record data for events appearing during a 10-ns-wide window synchronized to the laser pulse.  Note that the relative phases of the interfering pulses are not randomized on a pulse-to-pulse basis, but are on the scale of our total measurement time (typically 5 s).

\begin{figure}[tbh]
   \centering
     \resizebox{\columnwidth}{!}{\includegraphics{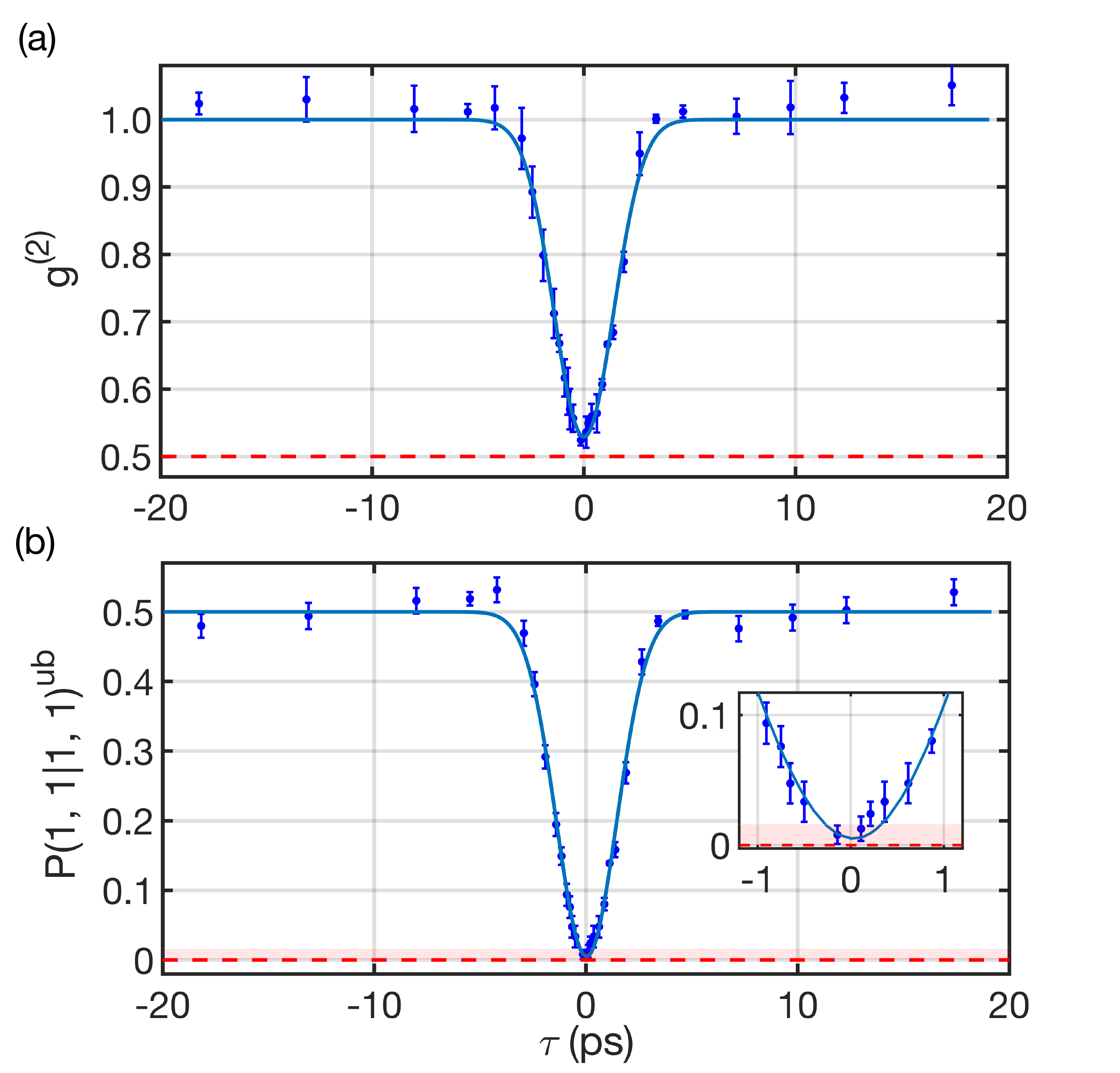}}
    \caption{(a) Second-order correlation function. (b) Observation of two-photon interference using PRWCS. The laser pulse repetition rate is 31.25 MHz, $\kappa\mu=(3.15\pm0.08)\times 10^{-3}$, and we collect four data sets at each delay with a 5-s-long counting interval for each. Vertical error bars correspond to the standard deviation of the four trials.
    \label{figure_g2}}
\end{figure}

Figure \ref{figure_g2}(a) shows the normalized coincidence count probability $C^{\mu,\mu}/S^1S^2=g^{(2)}(\tau)$ for equal and low mean photon numbers as the optical delay $\tau$ is adjusted, where $g^{(2)}$ is normalized intensity-intensity (second-order) correlation function.  As $\tau$ approaches zero from either side, $g^{(2)}$ decreases with a characteristic shape related to the auto-correlation of the wavepacket temporal profile.  We fit the data with an inverted and offset Gaussian function that takes on the value of 1 for large delay, and the width, temporal offset, and minimum value are left as fit parameters. At zero delay, $g^{(2)}(0)$=$0.529 \pm 0.015$ from the fit-function minima as expected based on the discussion above, where the errors represent the 95\% confidence interval.  Our results are comparable to  the best obtained in previous experiments measuring photon-photon correlations with PRWCS \cite{2016_OE_Shields,2016_NatPhot_Shields,2012_PRL_Lo, 2013_Nat_Comm_Jeongwan,2016_CPB_Chen,2018-Bing}.

We use Eq.~\ref{P11bound-cnts} to bound the outcome of the two-photon interference experiment as shown in Fig. \ref{figure_g2}(b). For large relative delay, $P(1,1|1,1)^{ub}\sim 0.5$ as expected for distinguishable photons. As the delay approaches zero from either side, $P(1,1|1,1)^{ub}$ decreases.  We fit the data with an inverted and offset Gaussian function that takes on the value of 1/2 for large delay, and the width, temporal offset, and minimum value are left as fit parameters.

We find $P(1,1|1,1)^{ub}=0.005^{+0.013}_{-0.005}$ at $\tau=0$, clearly revealing the HOM two-photon interference effect.  Here, the positive error represent the 95\% confidence interval of the fit.  The negative confidence interval (-0.008) places the bound less then zero, which is not physical for a probability, so we set this error to make the bound consistent with zero. Quantum theory predicts $P(1,1|1,1)^q=(R-T)^2=0.0016$ using the nominal values for $R$ and $T$, although it could range between 0 and 0.0064. The semi-transparent band in the figure indicates this possible range. Our experimental observations are consistent with this prediction to within our measurement errors.

In terms of the conditional probability, the visibility can be written as 
\begin{equation}
V=1-\frac{P(1,1|1,1)^{ub}}{2P(1,1|1,1)^c},
\end{equation}
where $P(1,1|1,1)^c=1/2$ is the predicted value for classical particles \cite{2017_Arxiv_Curty}.  We find that $V=0.995^{+0.005}_{-0.013}$, which is comparable to the best of any experimental observation.  This result is made possible by the ease of generating highly indistinguishable PRWCS using attenuated laser light.

\begin{figure}[tbh]
  \centering
    \resizebox{\columnwidth}{!}{\includegraphics{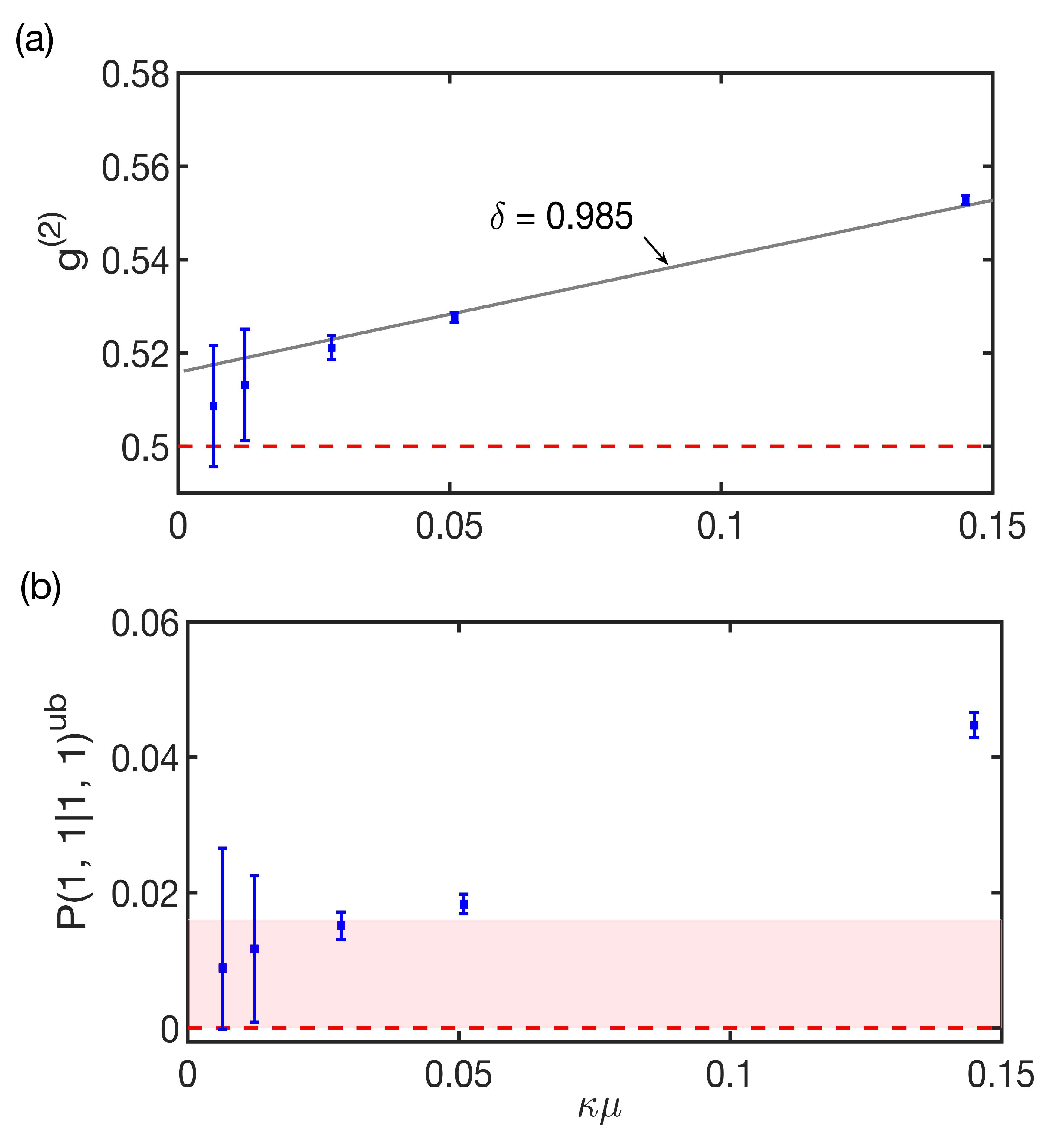}}
    \caption{Dependence of two-photon interference on mean photon number. (a) Intensity-intensity correlation function at $\tau=0$ and (b) the upper bound of the conditional probability related to two-photon interference. The horizontal dashed line at $g^{(2)}=0.5$ in (a) corresponds to the value for an ideal setup with $\mu \rightarrow 0$, while the solid line corresponds to the predictions of quantum theory accounting for the setup non-idealities. \label{figure_mu}}
 \end{figure}

As $\mu$ increases, the measured values of $g^{(2)}(0)$ increases as shown Fig. \ref{figure_mu}(a).  Here, we adjust $\mu$ by changing the coupling of the laser light into the single-mode fiber so that there is no misalignment of the beams at the beam splitter where interference takes place. To compare to predictions based on quantum theory, we use Eqs. (11)(15) of Ref. \cite{2018-Bing}.  We include the beamsplitter characteristics, assume $\mu_a=\mu_b=\mu$ and $\kappa_1=\kappa_2=\kappa$ (the predictions are rather insensitive to these assumptions).  Furthermore, we ignore detector dark counts and afterpulsing, which is appropriate for our low-noise detectors.

We find reasonable agreement between our observations and theoretical predictions ($\chi_R^2=0.92$) when we use an indistinguishabilty parameter $\delta=0.985$, where $\delta=1$ (0) when the pulses are perfectly indistinguishable (distinguishable)~\footnote{$\delta=\cos\Phi$ in the notation of Ref. \cite{2018-Bing}}. This small non-ideality may be due to tiny angular or spatial misalignment between the interfering beams due to creep of the translation stage in our setup or from other mechanical instabilities.

Even with these slight imperfections, we find that $P(1,1|1,1)^{ub}$ remains deep within the quantum regime (\textit{i.e.}, $<$1/2) over the entire range of mean photon numbers, as shown in Fig. \ref{figure_mu}(b). The red dashed line is the expected value based on quantum theory for an ideal setup and the semi-transparent band shows the range of values predicted for our non-ideal setup.  For the largest value of $\mu$, the measurement falls outside this band. This is likely due to the higher-order terms ignored when deriving the upper bound.  The accuracy of the bound can be improved using additional states with intermediate photon numbers \cite{Yuan-2016,2017_Arxiv_Curty}, which we will explore in future work.  However, even with this increase, $P(1,1|1,1)^{ub}\ll 1/2$.

Another interpretation of our work is that $P(1,1|1,1)^{ub}$ provides a sensitive measure of the indistinguishability of the single-photon wavepackets \cite{2018-Armal}, which has application to quantum key distribution.  In particular, an eavesdropper will necessarily disturb the quantum photonic wavepacket that have only a single photon, which can be detected by mixing the received wavepacket with an identical local-oscillator wavepacket and using the bounding technique described here.

Our approach may also find application in other experiments that apparently required the use of single-photon wavepackets.  As discussed by Yuan \textit{et al.} and Navarrete \textit{et al.}, it is possible to put tight upper and lower bounds on the outcome of quantum experiments for a wide range of linear photonic circuits using PRWCS with various values of the mean photon numbers and truncation of the corresponding higher-order terms. It should also be possible to apply our general approach to other states such as thermal light.

Finally, the photon fusion shown in Fig. \ref{fig1}(b) and (c) represents a highly entangled, two-photon N00N state, which can be used for enhanced metrology \cite{2008-Dowling}.  The output ports of the beam splitter can be directed to other optical systems, such as an interferometer, for increased sensitivity to changes in the phase of an object, for example.  Of course, additional measurements using states with varying mean photon numbers are required in this approach, but no post-selection of events is required. Further analysis is needed to determine whether this approach offers any advantage, which we will report on elsewhere.

\section{FUNDING INFORMATION}

Supported by the Office of Naval Research MURI program on Wavelength-Agile Quantum Key Distribution in a Marine Environment, Grant \# N00014-13-1-0627. 

\section{ACKNOWLEDGEMENTS}

We gratefully acknowledge discussions of this work with Marcos Curty, Hoi-Kwong Lo, Norbert L\"utkenhaus, \'{A}lvaro Navarrete, and Bing Qi. 

% Bibliography
\bibliography{AragonesesHOM.bib}

\begin{thebibliography}{18}%
\makeatletter
\providecommand \@ifxundefined [1]{%
 \@ifx{#1\undefined}
}%
\providecommand \@ifnum [1]{%
 \ifnum #1\expandafter \@firstoftwo
 \else \expandafter \@secondoftwo
 \fi
}%
\providecommand \@ifx [1]{%
 \ifx #1\expandafter \@firstoftwo
 \else \expandafter \@secondoftwo
 \fi
}%
\providecommand \natexlab [1]{#1}%
\providecommand \enquote  [1]{``#1''}%
\providecommand \bibnamefont  [1]{#1}%
\providecommand \bibfnamefont [1]{#1}%
\providecommand \citenamefont [1]{#1}%
\providecommand \href@noop [0]{\@secondoftwo}%
\providecommand \href [0]{\begingroup \@sanitize@url \@href}%
\providecommand \@href[1]{\@@startlink{#1}\@@href}%
\providecommand \@@href[1]{\endgroup#1\@@endlink}%
\providecommand \@sanitize@url [0]{\catcode `\\12\catcode `\$12\catcode
  `\&12\catcode `\#12\catcode `\^12\catcode `\_12\catcode `\%12\relax}%
\providecommand \@@startlink[1]{}%
\providecommand \@@endlink[0]{}%
\providecommand \url  [0]{\begingroup\@sanitize@url \@url }%
\providecommand \@url [1]{\endgroup\@href {#1}{\urlprefix }}%
\providecommand \urlprefix  [0]{URL }%
\providecommand \Eprint [0]{\href }%
\providecommand \doibase [0]{http://dx.doi.org/}%
\providecommand \selectlanguage [0]{\@gobble}%
\providecommand \bibinfo  [0]{\@secondoftwo}%
\providecommand \bibfield  [0]{\@secondoftwo}%
\providecommand \translation [1]{[#1]}%
\providecommand \BibitemOpen [0]{}%
\providecommand \bibitemStop [0]{}%
\providecommand \bibitemNoStop [0]{.\EOS\space}%
\providecommand \EOS [0]{\spacefactor3000\relax}%
\providecommand \BibitemShut  [1]{\csname bibitem#1\endcsname}%
\let\auto@bib@innerbib\@empty
%</preamble>
\bibitem [{\citenamefont {Hong}\ \emph {et~al.}(1987)\citenamefont {Hong},
  \citenamefont {Ou},\ and\ \citenamefont {Mandel}}]{1987_PRL_HOM}%
  \BibitemOpen
  \bibfield  {author} {\bibinfo {author} {\bibfnamefont {C.~K.}\ \bibnamefont
  {Hong}}, \bibinfo {author} {\bibfnamefont {Z.~Y.}\ \bibnamefont {Ou}}, \ and\
  \bibinfo {author} {\bibfnamefont {L.}~\bibnamefont {Mandel}},\ }\href
  {\doibase 10.1103/PhysRevLett.59.2044} {\bibfield  {journal} {\bibinfo
  {journal} {Phys. Rev. Lett.}\ }\textbf {\bibinfo {volume} {59}},\ \bibinfo
  {pages} {2044} (\bibinfo {year} {1987})}\BibitemShut {NoStop}%
\bibitem [{\citenamefont {Walmsley}(2017)}]{Walmsley-2017}%
  \BibitemOpen
  \bibfield  {author} {\bibinfo {author} {\bibfnamefont {I.}~\bibnamefont
  {Walmsley}},\ }\href {\doibase 10.1126/science.aao3883} {\bibfield  {journal}
  {\bibinfo  {journal} {Science}\ }\textbf {\bibinfo {volume} {258}},\ \bibinfo
  {pages} {1001} (\bibinfo {year} {2017})}\BibitemShut {NoStop}%
\bibitem [{\citenamefont {Mandel}(1983)}]{1983_PRA_Mandel}%
  \BibitemOpen
  \bibfield  {author} {\bibinfo {author} {\bibfnamefont {L.}~\bibnamefont
  {Mandel}},\ }\href {\doibase 10.1103/PhysRevA.28.929} {\bibfield  {journal}
  {\bibinfo  {journal} {Phys. Rev. A}\ }\textbf {\bibinfo {volume} {28}},\
  \bibinfo {pages} {929} (\bibinfo {year} {1983})}\BibitemShut {NoStop}%
\bibitem [{\citenamefont {Rarity}\ \emph {et~al.}(2005)\citenamefont {Rarity},
  \citenamefont {Tapster},\ and\ \citenamefont {Loudon}}]{2005_JOB_Loudon}%
  \BibitemOpen
  \bibfield  {author} {\bibinfo {author} {\bibfnamefont {J.~G.}\ \bibnamefont
  {Rarity}}, \bibinfo {author} {\bibfnamefont {P.~R.}\ \bibnamefont {Tapster}},
  \ and\ \bibinfo {author} {\bibfnamefont {R.}~\bibnamefont {Loudon}},\ }\href
  {http://stacks.iop.org/1464-4266/7/i=7/a=007} {\bibfield  {journal} {\bibinfo
   {journal} {Journal of Optics B: Quantum and Semiclassical Optics}\ }\textbf
  {\bibinfo {volume} {7}},\ \bibinfo {pages} {S171} (\bibinfo {year}
  {2005})}\BibitemShut {NoStop}%
\bibitem [{\citenamefont {Jin}\ \emph {et~al.}(2013)\citenamefont {Jin},
  \citenamefont {Slater}, \citenamefont {Saglamyurek}, \citenamefont
  {Sinclair}, \citenamefont {George}, \citenamefont {Ricken}, \citenamefont
  {Oblak}, \citenamefont {Sohler},\ and\ \citenamefont
  {Tittel}}]{2013_Nat_Comm_Jeongwan}%
  \BibitemOpen
  \bibfield  {author} {\bibinfo {author} {\bibfnamefont {J.}~\bibnamefont
  {Jin}}, \bibinfo {author} {\bibfnamefont {J.~A.}\ \bibnamefont {Slater}},
  \bibinfo {author} {\bibfnamefont {E.}~\bibnamefont {Saglamyurek}}, \bibinfo
  {author} {\bibfnamefont {N.}~\bibnamefont {Sinclair}}, \bibinfo {author}
  {\bibfnamefont {M.}~\bibnamefont {George}}, \bibinfo {author} {\bibfnamefont
  {R.}~\bibnamefont {Ricken}}, \bibinfo {author} {\bibfnamefont
  {D.}~\bibnamefont {Oblak}}, \bibinfo {author} {\bibfnamefont
  {W.}~\bibnamefont {Sohler}}, \ and\ \bibinfo {author} {\bibfnamefont
  {W.}~\bibnamefont {Tittel}},\ }\href@noop {} {\bibfield  {journal} {\bibinfo
  {journal} {Nature Communications}\ }\textbf {\bibinfo {volume} {4}},\
  \bibinfo {pages} {2386} (\bibinfo {year} {2013})}\BibitemShut {NoStop}%
\bibitem [{\citenamefont {Chen}\ \emph {et~al.}(2016)\citenamefont {Chen},
  \citenamefont {An}, \citenamefont {Wu}, \citenamefont {Yin}, \citenamefont
  {Wang}, \citenamefont {Chen},\ and\ \citenamefont {Han}}]{2016_CPB_Chen}%
  \BibitemOpen
  \bibfield  {author} {\bibinfo {author} {\bibfnamefont {H.}~\bibnamefont
  {Chen}}, \bibinfo {author} {\bibfnamefont {X.-B.}\ \bibnamefont {An}},
  \bibinfo {author} {\bibfnamefont {J.}~\bibnamefont {Wu}}, \bibinfo {author}
  {\bibfnamefont {Z.-Q.}\ \bibnamefont {Yin}}, \bibinfo {author} {\bibfnamefont
  {S.}~\bibnamefont {Wang}}, \bibinfo {author} {\bibfnamefont {W.}~\bibnamefont
  {Chen}}, \ and\ \bibinfo {author} {\bibfnamefont {Z.-F.}\ \bibnamefont
  {Han}},\ }\href@noop {} {\bibfield  {journal} {\bibinfo  {journal} {Chin.
  Phys. B}\ }\textbf {\bibinfo {volume} {25}},\ \bibinfo {pages} {020305}
  (\bibinfo {year} {2016})}\BibitemShut {NoStop}%
\bibitem [{\citenamefont {Valivarthi}\ \emph {et~al.}(2017)\citenamefont
  {Valivarthi}, \citenamefont {Zhou}, \citenamefont {Caleb}, \citenamefont
  {Marsili}, \citenamefont {Verma}, \citenamefont {Shaw}, \citenamefont {Nam},
  \citenamefont {Oblak},\ and\ \citenamefont {Tittel}}]{2017_QST_Tittel}%
  \BibitemOpen
  \bibfield  {author} {\bibinfo {author} {\bibfnamefont {R.}~\bibnamefont
  {Valivarthi}}, \bibinfo {author} {\bibfnamefont {Q.}~\bibnamefont {Zhou}},
  \bibinfo {author} {\bibfnamefont {J.}~\bibnamefont {Caleb}}, \bibinfo
  {author} {\bibfnamefont {F.}~\bibnamefont {Marsili}}, \bibinfo {author}
  {\bibfnamefont {V.~B.}\ \bibnamefont {Verma}}, \bibinfo {author}
  {\bibfnamefont {M.~D.}\ \bibnamefont {Shaw}}, \bibinfo {author}
  {\bibfnamefont {S.~W.}\ \bibnamefont {Nam}}, \bibinfo {author} {\bibfnamefont
  {D.}~\bibnamefont {Oblak}}, \ and\ \bibinfo {author} {\bibfnamefont
  {W.}~\bibnamefont {Tittel}},\ }\href@noop {} {\bibfield  {journal} {\bibinfo
  {journal} {Quant. Scie. Tech}\ }\textbf {\bibinfo {volume} {2}},\ \bibinfo
  {pages} {04LT01} (\bibinfo {year} {2017})}\BibitemShut {NoStop}%
\bibitem [{\citenamefont {Moschandreou}\ \emph {et~al.}(2018)\citenamefont
  {Moschandreou}, \citenamefont {Garcia}, \citenamefont {Rollick},
  \citenamefont {Qi}, \citenamefont {Pooser},\ and\ \citenamefont
  {Siopsis}}]{2018-Bing}%
  \BibitemOpen
  \bibfield  {author} {\bibinfo {author} {\bibfnamefont {E.}~\bibnamefont
  {Moschandreou}}, \bibinfo {author} {\bibfnamefont {J.~I.}\ \bibnamefont
  {Garcia}}, \bibinfo {author} {\bibfnamefont {B.~J.}\ \bibnamefont {Rollick}},
  \bibinfo {author} {\bibfnamefont {B.}~\bibnamefont {Qi}}, \bibinfo {author}
  {\bibfnamefont {R.}~\bibnamefont {Pooser}}, \ and\ \bibinfo {author}
  {\bibfnamefont {G.}~\bibnamefont {Siopsis}},\ }\href@noop {} {\bibfield
  {journal} {\bibinfo  {journal} {arXiv:1804.0229}\ } (\bibinfo {year}
  {2018})}\BibitemShut {NoStop}%
\bibitem [{\citenamefont {Yuan}\ \emph {et~al.}(2016)\citenamefont {Yuan},
  \citenamefont {Zhang}, \citenamefont {L\"utkenhaus},\ and\ \citenamefont
  {Ma}}]{Yuan-2016}%
  \BibitemOpen
  \bibfield  {author} {\bibinfo {author} {\bibfnamefont {X.}~\bibnamefont
  {Yuan}}, \bibinfo {author} {\bibfnamefont {Z.}~\bibnamefont {Zhang}},
  \bibinfo {author} {\bibfnamefont {N.}~\bibnamefont {L\"utkenhaus}}, \ and\
  \bibinfo {author} {\bibfnamefont {X.}~\bibnamefont {Ma}},\ }\href {\doibase
  10.1103/PhysRevA.94.062305} {\bibfield  {journal} {\bibinfo  {journal} {Phys.
  Rev. A}\ }\textbf {\bibinfo {volume} {94}},\ \bibinfo {pages} {062305}
  (\bibinfo {year} {2016})}\BibitemShut {NoStop}%
\bibitem [{\citenamefont {Navarrete}\ \emph {et~al.}(2018)\citenamefont
  {Navarrete}, \citenamefont {Wang}, \citenamefont {Xu},\ and\ \citenamefont
  {Curty}}]{2017_Arxiv_Curty}%
  \BibitemOpen
  \bibfield  {author} {\bibinfo {author} {\bibfnamefont {{\'A}.}~\bibnamefont
  {Navarrete}}, \bibinfo {author} {\bibfnamefont {W.}~\bibnamefont {Wang}},
  \bibinfo {author} {\bibfnamefont {F.}~\bibnamefont {Xu}}, \ and\ \bibinfo
  {author} {\bibfnamefont {M.}~\bibnamefont {Curty}},\ }\href@noop {}
  {\bibfield  {journal} {\bibinfo  {journal} {New J. Phys.}\ }\textbf {\bibinfo
  {volume} {20}},\ \bibinfo {pages} {043018} (\bibinfo {year}
  {2018})}\BibitemShut {NoStop}%
\bibitem [{\citenamefont {Valente}\ and\ \citenamefont
  {Lezama}(2017)}]{2017_JOSAB_Lezama}%
  \BibitemOpen
  \bibfield  {author} {\bibinfo {author} {\bibfnamefont {P.}~\bibnamefont
  {Valente}}\ and\ \bibinfo {author} {\bibfnamefont {A.}~\bibnamefont
  {Lezama}},\ }\href {\doibase 10.1364/JOSAB.34.000924} {\bibfield  {journal}
  {\bibinfo  {journal} {J. Opt. Soc. Am. B}\ }\textbf {\bibinfo {volume}
  {34}},\ \bibinfo {pages} {924} (\bibinfo {year} {2017})}\BibitemShut
  {NoStop}%
\bibitem [{Note1()}]{Note1}%
  \BibitemOpen
  \bibinfo {note} {In the quantum cryptography community, the largest values of
  $\mu _a$ and $\mu _b$ are often called `signal' states, whereas smaller
  values (including zero mean photon number) are often called `decoy'
  states.}\BibitemShut {Stop}%
\bibitem [{\citenamefont {Comandar}\ \emph
  {et~al.}(2016{\natexlab{a}})\citenamefont {Comandar}, \citenamefont
  {Lucamarini}, \citenamefont {Fr\"{o}hlich}, \citenamefont {Dynes},
  \citenamefont {Yuan},\ and\ \citenamefont {Shields}}]{2016_OE_Shields}%
  \BibitemOpen
  \bibfield  {author} {\bibinfo {author} {\bibfnamefont {L.~C.}\ \bibnamefont
  {Comandar}}, \bibinfo {author} {\bibfnamefont {M.}~\bibnamefont
  {Lucamarini}}, \bibinfo {author} {\bibfnamefont {B.}~\bibnamefont
  {Fr\"{o}hlich}}, \bibinfo {author} {\bibfnamefont {J.~F.}\ \bibnamefont
  {Dynes}}, \bibinfo {author} {\bibfnamefont {Z.~L.}\ \bibnamefont {Yuan}}, \
  and\ \bibinfo {author} {\bibfnamefont {A.~J.}\ \bibnamefont {Shields}},\
  }\href {\doibase 10.1364/OE.24.017849} {\bibfield  {journal} {\bibinfo
  {journal} {Opt. Express}\ }\textbf {\bibinfo {volume} {24}},\ \bibinfo
  {pages} {17849} (\bibinfo {year} {2016}{\natexlab{a}})}\BibitemShut {NoStop}%
\bibitem [{\citenamefont {Comandar}\ \emph
  {et~al.}(2016{\natexlab{b}})\citenamefont {Comandar}, \citenamefont
  {Lucamarini}, \citenamefont {Fr{\"o}hlich}, \citenamefont {Dynes},
  \citenamefont {Sharpe}, \citenamefont {Tam}, \citenamefont {Yuan},
  \citenamefont {Penty},\ and\ \citenamefont {Shields}}]{2016_NatPhot_Shields}%
  \BibitemOpen
  \bibfield  {author} {\bibinfo {author} {\bibfnamefont {L.}~\bibnamefont
  {Comandar}}, \bibinfo {author} {\bibfnamefont {M.}~\bibnamefont
  {Lucamarini}}, \bibinfo {author} {\bibfnamefont {B.}~\bibnamefont
  {Fr{\"o}hlich}}, \bibinfo {author} {\bibfnamefont {J.}~\bibnamefont {Dynes}},
  \bibinfo {author} {\bibfnamefont {A.}~\bibnamefont {Sharpe}}, \bibinfo
  {author} {\bibfnamefont {S.-B.}\ \bibnamefont {Tam}}, \bibinfo {author}
  {\bibfnamefont {Z.}~\bibnamefont {Yuan}}, \bibinfo {author} {\bibfnamefont
  {R.}~\bibnamefont {Penty}}, \ and\ \bibinfo {author} {\bibfnamefont
  {A.}~\bibnamefont {Shields}},\ }\href@noop {} {\bibfield  {journal} {\bibinfo
   {journal} {Nature Photonics}\ }\textbf {\bibinfo {volume} {10}},\ \bibinfo
  {pages} {312} (\bibinfo {year} {2016}{\natexlab{b}})}\BibitemShut {NoStop}%
\bibitem [{\citenamefont {Lo}\ \emph {et~al.}(2012)\citenamefont {Lo},
  \citenamefont {Curty},\ and\ \citenamefont {Qi}}]{2012_PRL_Lo}%
  \BibitemOpen
  \bibfield  {author} {\bibinfo {author} {\bibfnamefont {H.-K.}\ \bibnamefont
  {Lo}}, \bibinfo {author} {\bibfnamefont {M.}~\bibnamefont {Curty}}, \ and\
  \bibinfo {author} {\bibfnamefont {B.}~\bibnamefont {Qi}},\ }\href {\doibase
  10.1103/PhysRevLett.108.130503} {\bibfield  {journal} {\bibinfo  {journal}
  {Phys. Rev. Lett.}\ }\textbf {\bibinfo {volume} {108}},\ \bibinfo {pages}
  {130503} (\bibinfo {year} {2012})}\BibitemShut {NoStop}%
\bibitem [{Note2()}]{Note2}%
  \BibitemOpen
  \bibinfo {note} {$\delta =\protect \qopname \relax o{cos}\Phi $ in the
  notation of Ref. \cite {2018-Bing}}\BibitemShut {NoStop}%
\bibitem [{\citenamefont {Armal}\ \emph {et~al.}(2018)\citenamefont {Armal},
  \citenamefont {Carneiro}, \citenamefont {Tempor{\~a}o},\ and\ \citenamefont
  {von~der Weid}}]{2018-Armal}%
  \BibitemOpen
  \bibfield  {author} {\bibinfo {author} {\bibfnamefont {G.~C.}\ \bibnamefont
  {Armal}}, \bibinfo {author} {\bibfnamefont {E.~F.}\ \bibnamefont {Carneiro}},
  \bibinfo {author} {\bibfnamefont {G.~P.}\ \bibnamefont {Tempor{\~a}o}}, \
  and\ \bibinfo {author} {\bibfnamefont {J.~P.}\ \bibnamefont {von~der Weid}},\
  }\href@noop {} {\bibfield  {journal} {\bibinfo  {journal} {J. Opt. Soc. Am.
  B}\ }\textbf {\bibinfo {volume} {35}},\ \bibinfo {pages} {601} (\bibinfo
  {year} {2018})}\BibitemShut {NoStop}%
\bibitem [{\citenamefont {Dowling}(2008)}]{2008-Dowling}%
  \BibitemOpen
  \bibfield  {author} {\bibinfo {author} {\bibfnamefont {J.~P.}\ \bibnamefont
  {Dowling}},\ }\href@noop {} {\bibfield  {journal} {\bibinfo  {journal}
  {Contemp. Phys.}\ }\textbf {\bibinfo {volume} {49}} (\bibinfo {year}
  {2008})}\BibitemShut {NoStop}%
\end{thebibliography}%

% Full bibliography added automatically for Optics Letters submissions; the following line will simply be ignored if submitting to other journals.
% Note that this extra page will not count against page length
%\bibliographyfullrefs{AragonesesHOM}

\end{document}